\documentclass{elsart}
\usepackage{epsfig}
\begin{document}
\runauthor{Postnikov {\it et al.}}
\begin{frontmatter}
\title{Combined study of KNbO$_3$ and KTaO$_3$
by different techniques of photoelectron and
X-ray emission spectroscopy}
\author{A.~V.~Postnikov, B.~Schneider,
M.~Neumann, D.~Hartmann, H.~Hesse}
\address{
Universit\"at Osnabr\"uck -- Fachbereich Physik,\\
D-49069 Osnabr\"uck, Germany}
\author{A.~Moewes}
\address{
Center for Advanced Microstructures and Devices at
Louisiana State University, Baton Rouge, LA 70803, USA}
\author{E.~Z.~Kurmaev}
\address{
Institute of Metal Physics, GSP-170 Yekaterinburg, Russia}
\author{M.~Matteucci}
\address{
National Research Council,
c/o Sincrotrone Trieste, Padriciano 99, I-34012 Trieste, Italy}
\begin{abstract}
The new results are presented
of the experimental study of KNbO$_3$ and KTaO$_3$
by means of X-ray photoelectron spectroscopy,
soft X-ray emission and X-ray fluorescence spectroscopy.
In particular, Nb $M_{4,5}$ spectra
($5p\,4f\!\rightarrow\!3d_{5/2,\,3/2}$ transition)
have been measured over a range of
near-threshold excitation energies
206.5 -- 240.6 eV,
and Ta $N_3$ spectra ($5d\,6s\!\rightarrow\!4p_{3/2}$ transition)
over 399.4 -- 420.4 eV at the Beamline 8.0
of the Lawrence Berkeley National Laboratory's
Advance Light Source.
These spectra were found to be strongly dependent
on the excitation energy.
Moreover, the O $K_{\alpha}$ X-ray emission
spectra in both compounds as well as the valence-band photoelectron
spectra are brought to the binding energy scale and discussed
in combination.
The trends in the spectra are explained on the basis
of first-principles band-structure calculations,
with the dipole transition matrix elements taken into account.
\end{abstract}
\end{frontmatter}

\section*{Introduction}
Potassium niobate and tantalate are traditional
benchmark systems for the testing of new theory developments
in the study of ferroelectric materials.
The starting point of any first-principles treatment,
aimed at lattice dynamics or dielectric response,
is the knowledge of the ground-state electronic structure.
That of KNbO$_3$ and KTaO$_3$ is believed
to be quite well known as a result of
a long development beginning with the empirical parameter-adjusting
schemes by the end of 1970s \cite{pertosa}, followed by
first-principles self-consistent treatment \cite{opw,asa92}
and finally refined in precision all-electron total energy
calculations by different methods \cite{sb92,ktn3,ksv94}.
In the evaluation of one-electron Kohn-Sham energies,
an excellent agreement exists nowadays between different
technical implementations of the density functional theory (DFT) --
see, for example, Ref. \cite{ktn3,ksv94,singh}.
However, the Hartree-Fock formalism
provides basically  different description of the
one-particle excitation spectrum and hence somehow different
dielectric constant than that in the DFT treatment \cite{HF-eps}.
In the situation when {\it ab initio} calculations of
quasiparticle excitations spectra for such moderately complex
systems as cubic perovskites are not yet routinely feasible,
the electronic and X-ray spectroscopy remains the important tool
for the experimental evaluation of the electronic structure data.
The X-ray photoelectron spectroscopy (XPS)
has been applied in several cases
to probe the density of states (DOS) distribution in the valence band
(VB) of KNbO$_3$ and KTaO$_3$ \cite{asa92,winiar,douill}.
Moreover, the angle-resolved ultraviolet photoelectron spectra
have been measured for KTaO$_3$ and explained in terms
of full-potential relativistic theory of photoemission \cite{ups96}.
While being often quite successful in the study of metals,
the electron spectroscopy techniques face severe problems in dielectrics
due to sample-charging effects. This problem does not arise
in the X-ray emission spectroscopy (XES). Larger exit depth
of soft X-ray emission, as compared to that of photoelectrons,
may be advantageous in many cases because the quality of the
sample surface preparation is not so crucial.
Moreover, due to dipole selection rules the XES
is element-sensitive, which allows to probe partial
states distribution related to different atoms.
It has been shown that not only DOS but also
momentum-resolved information regarding
the VB dispersion can be extracted from
soft X-ray resonant inelastic spectra (XRIS) \cite{xris}.
Therefore, XES complements the electron spectroscopy
in the study of electronic structure of insulating materials.
In the present work, we concentrate on
the partial DOS demonstrated by several X-ray emission
spectra in KNbO$_3$ and KTaO$_3$ and analyzed in comparison
with the VB XPS. The interpretation of spectra
(including the dipole transition matrix elements) was done
on the basis of calculations by the all-electron
full-potential linearized augmented plane-wave method (FLAPW).

\section*{Experiment and calculation details}
The soft X-ray fluorescence experiments
were performed at Beamline 8.0 of the Advanced
Light Source at Lawrence Berkeley Laboratory.
The undulator beamline is equipped with a
spherical grating monochromator \cite{monocrom}
and the resolving power was set to $E/{\Delta}E$=500.
The fluorescence end station with a Rowland circle
grating spectrometer provided a resolving power
of about 300 at 200 eV.
The XPS measurements have been done using a Perkin Elmer PHI 5600ci
Multitechnique System with monochromatized
Al $K\alpha$ radiation (bandwidth 0.3 eV FWHM).
In order to obtain clean surfaces for the XPS measurements,
the KNbO$_3$ single crystal was cleaved
under UHV conditions.
The X-ray Nb $L\beta_{2,15}$ and Nb $L\gamma_1$ fluorescence spectra
were obtained by a Stearat spectrometer \cite{stearat}.
A quartz crystal ($d$=0.334 nm) curved to $R$=500 cm was used as
dispersive element to analyze the photons.
X-ray emission spectra were detected by a flow
proportional counter by scanning along the Rowland circle
with an energy resolution of $\pm$0.2 eV.

The FLAPW calculations have been done with the WIEN97 implementation
of the code \cite{wien97}. With its extended basis
of augmented plane waves, the FLAPW method allows a practical description
of the electronic states up to relatively high energies
in the conduction band, which is important for analyzing the trends
in the X-ray absorption and the resonance X-ray
emission intensities \cite{note}.
The experimental lattice constant a=7.553 a.u. was used, and the atomic
sphere radii were set to 1.95 a.u. (K), 1.85 a.u. (Nb) and 1.65 a.u. (O).
The lattice was assumed to be an ideal cubic perovskite
because the effect of ferroelectric distortion on the DOS is known to
be negligible considering the comparison
to experimentally broadened spectra.
The local density approximation for the exchange-correlation was used,
according to the prescription by Perdew and Wang \cite{pw92}.
The states in the VB have been treated semi-relativistically,
the core states fully relativistically. The DOS and emission spectra
were calculated using the tetrahedron method,
for the $12\!\times\!12\!\times\!12$ divisions
of the whole Brillouin zone. Whereas our calculated partial
and total DOS agree with previous calculations cited above,
the calculated X-ray emission (and absorption)
spectra, including the energy dependence
of the dipole transition matrix elements present, to our
knowledge, new information.
We utilize our calculated results to understand the trends in our
resonant X-ray emission spectra.

\section*{Results and discussion}

\begin{figure}
\centerline{
\epsfig{file=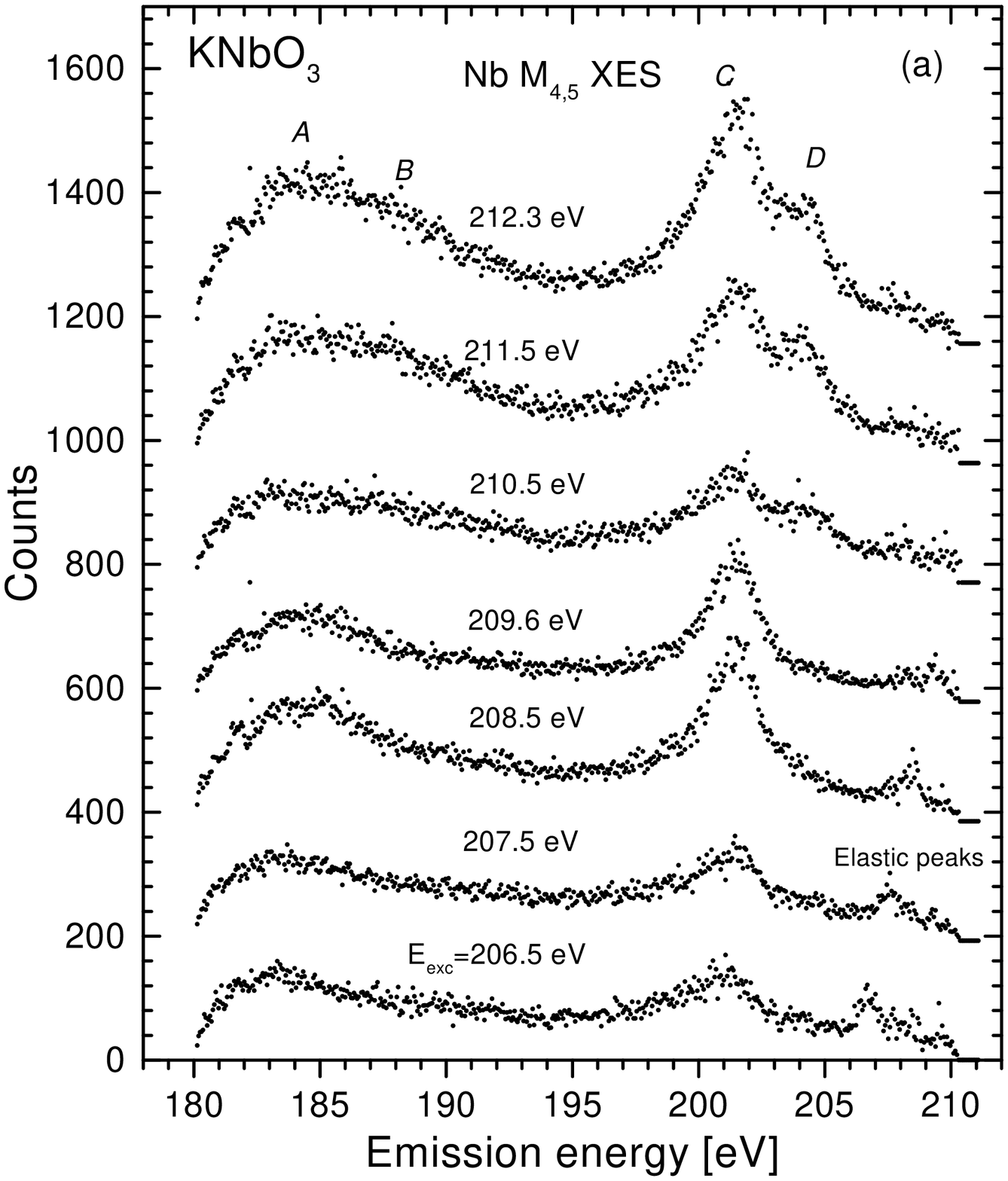,height=8.0cm}
\epsfig{file=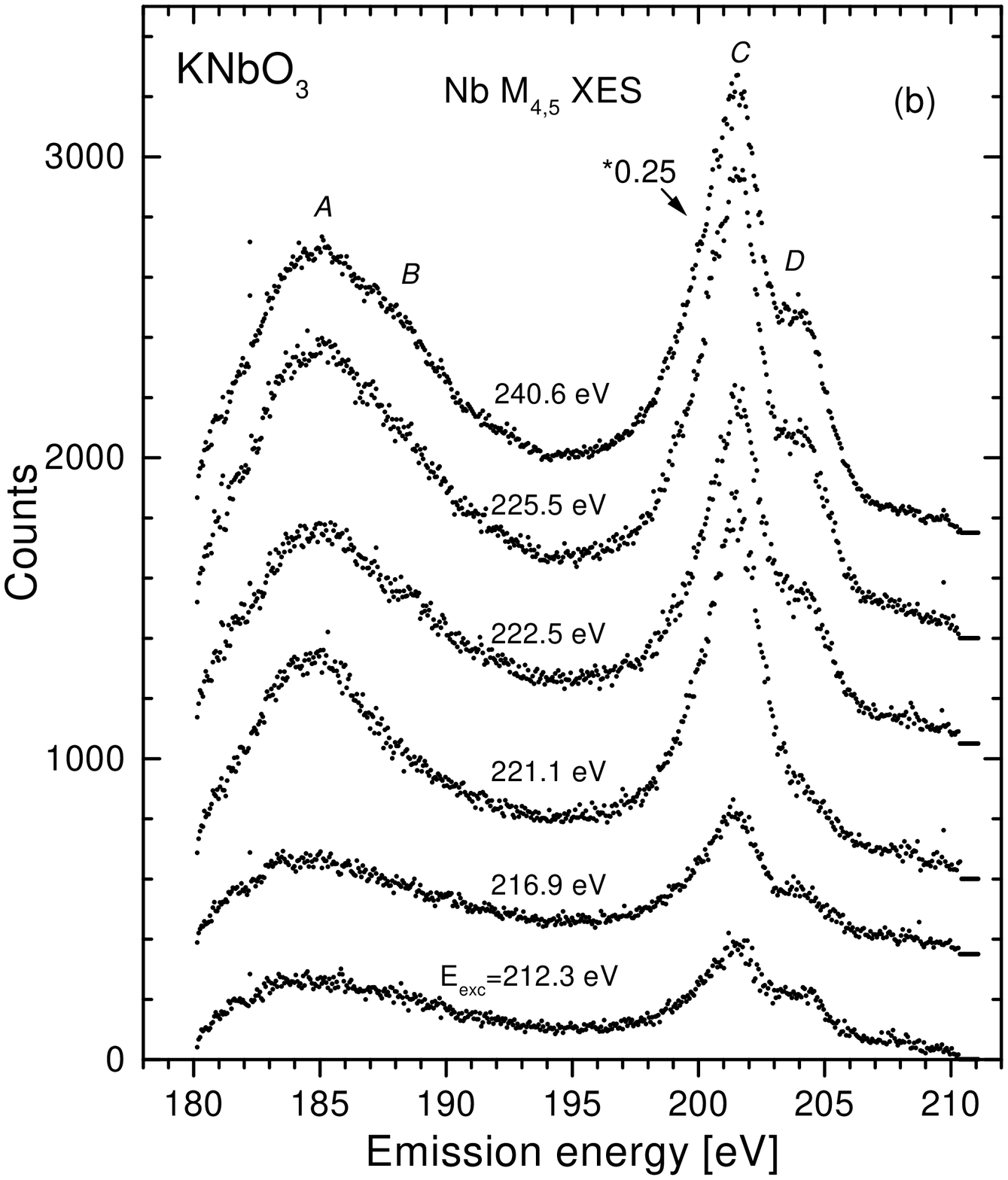,height=8.0cm}
}
\caption{
Nb $M_{4,5}$ emission spectra of KNbO$_3$.
The excitation energy is tuned through the Nb $3d$-threshold
from 206.5 to 212.3 eV (a) and from 212.3 to 240.6 eV (b).
}
\label{fig:M45-exp}
\end{figure}

In Fig.~\ref{fig:M45-exp} the Nb $M_{4,5}$ emission spectra of KNbO$_3$
are shown for various excitation
energies near the Nb$3d$ threshold. Four features are observed
and labeled $A$ through $D$. The most
dramatic changes in the fine structure of Nb $M_{4,5}$ emission spectra
are found for excitation energies between 206.5 and 212.3 eV
where new features $B$ and $D$ appear with changes in
excitation energy. It was suggested in Ref. \cite{lukir}
that the Nb $M_{4,5}$ emission only reveals the $M_5$ ($3d_{5/2}$)
features because the $M_4$ ($3d_{3/2}$) is filled
by radiationless transition. Therefore it was
unexpected when we found the excitation energy dependence
of Nb $M_{4,5}$ to be distorted by the
spin-orbit splitting of the Nb $3d$-levels and hence
the band dispersion effects to be blurred.

\begin{figure}
\centerline{\epsfig{file=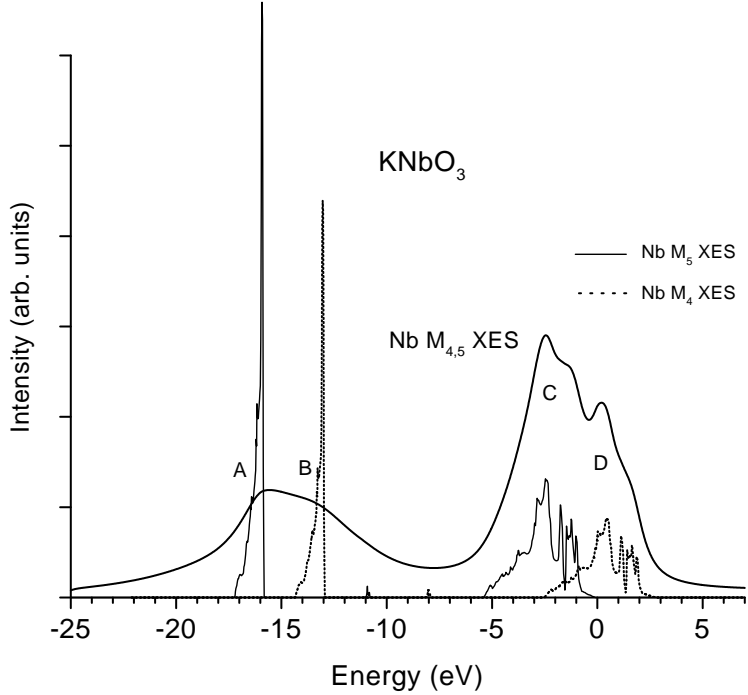,height=9.0cm}}
\caption{
Calculated Nb $M_{4,5}$ XES of KNbO$_3$ broadened for
instrumental broadening (0.2 eV), core level life-time (0.35 eV)
and valence life-time (2.0 eV). The contributions of unbroadened
$M_5$ and $M_4$ spectra are also shown.
}
\label{fig:M45-calc}
\end{figure}

Fig.~\ref{fig:M45-calc} displays the calculated
Nb $M_{4,5}$ emission spectra
based on the Nb$5p$ and Nb$4f$ DOS and modulated by the dipole
transition probabilities. The resulting spectrum consists
of two (identical) contributions, corresponding to individual
$M_4$ and $M_5$ spectra, that have been shifted apart by the value
of the calculated Nb$3d$ spin-orbit splitting (2.88 eV)
and summed up with relative weights 2:3.
It was found that the contribution of the Nb$4f$ states is
negligible in the VB, so that only the $5p{\rightarrow}3d$
transition is important for the interpretation of the Nb $M_{4,5}$ XES.
The O$2s$ states contribute to the Nb $M_5$ XES due to the O$2s-$Nb$5p$
hybridization. The corresponding features in the Nb $M_4$ and $M_5$ XES
near $-16$ eV and $-13$ eV are broadened in the
experimental spectra but still recognizable in Fig.~\ref{fig:M45-exp}
as peaks $A$ and $B$.

\begin{figure}
\centerline{\epsfig{file=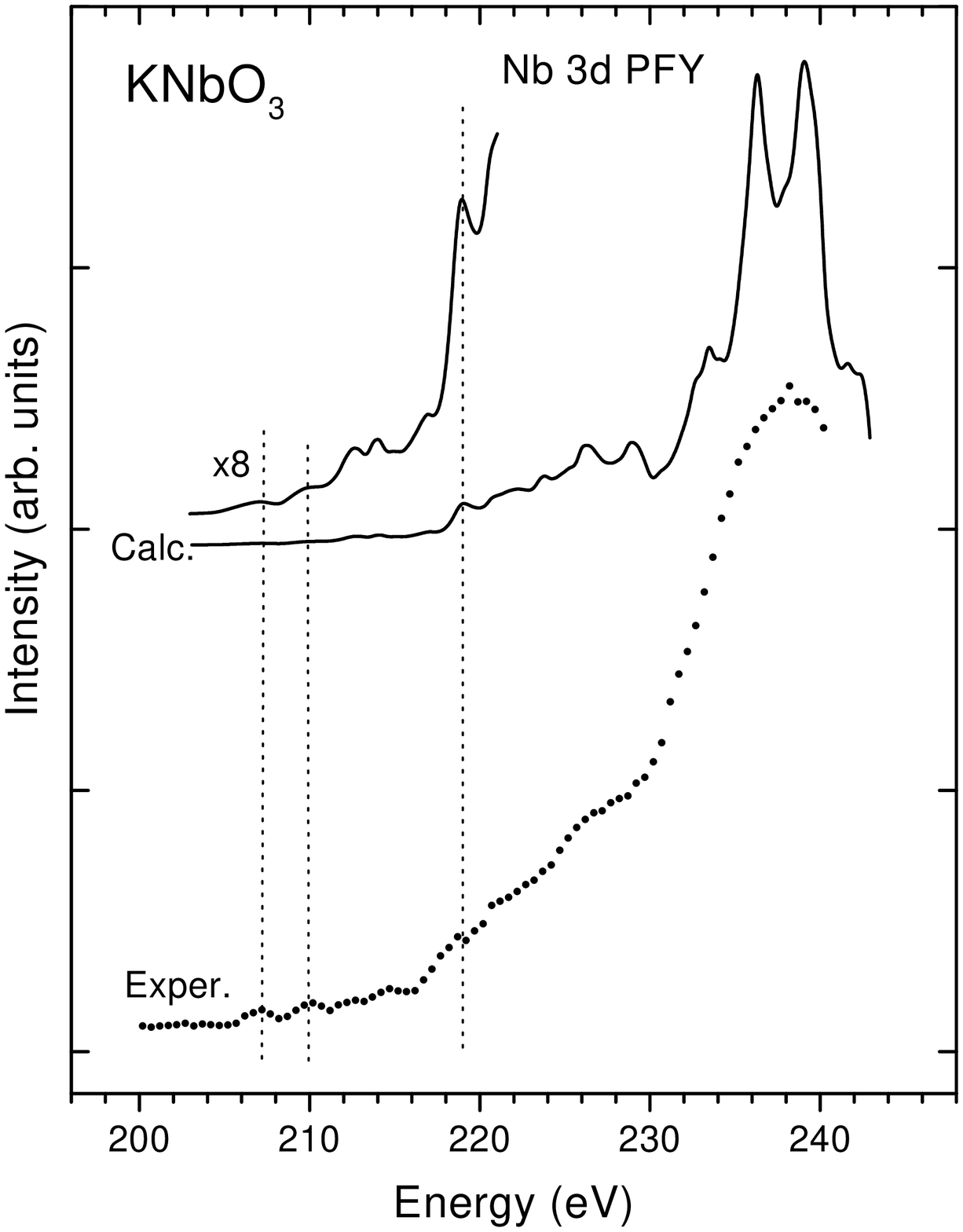,height=11.0cm}}
\caption{
Calculated and measured
Nb $3d{\rightarrow}5p$ partial fluorescence yield (PFY).
}
\label{fig:absor}
\end{figure}

Going back to the discussion of the energy dependence
of the Nb $M_{4,5}$ emission spectra (Fig.~\ref{fig:M45-exp}),
we note that according to our XPS measurements
the Nb $M_5$ ($3d_{5/2}$) and Nb $M_4$ ($3d_{3/2}$) binding energies
are 207.23 and 210 eV, respectively (relative to the vacuum level).
Conclusively the emission features in the spectra
excited between 206.5 and 209.6 eV are generated by the refill
of the Nb $M_5$($3d_{5/2}$) hole because the excitation of Nb $M_4$
is not possible below 210 eV. In the excitation energy range
between 210.5 and 216.9 eV, additional features $B$ and $D$ appear
as a result of contributing transitions to Nb $M_4$ ($3d_{3/2}$).
The sharp increase in intensity in the emission spectra
at excitation energies from 216.9 eV to 221.1 eV can be
attributed to the threshold of the $3d{\rightarrow}4f$ absorption.
This is illustrated by the calculated Nb $M_4$ absorption spectrum
in Fig.~\ref{fig:absor}. The onset for the $3d{\rightarrow}4f$ absorption
occurs around 235 eV. Whereas the $3d{\rightarrow}4f$ contribution
dominates in the $M_{4,5}$ absorption spectrum, the
$3d{\rightarrow}5p$ process gives rise to the absorption
at about 218 eV. The strong enhancement of the
emission (see Fig.~\ref{fig:M45-exp}) occurs
when the excitation energy exceeds
the $M_5$ threshold ($E_{exc}$=208.5 eV),
$M_4$ threshold ($E_{exc}$=211.5 eV) and the
$3d-5p$ threshold ($E_{exc}$=221.1eV) as displayed in the
absorption spectrum (Fig.~\ref{fig:absor}).
This sharp increase in the absorption intensity
and hence in the resonant emission appears in each of these steps,
at increase of the excitation energy, first for the $M_5$ component
and followed by the $M_4$. The emission spectrum corresponding to
the excitation energy of 221.1 eV in Fig.~\ref{fig:M45-exp} exhibits,
as compared with that for $E_{exc}$=216.9 eV,
strongly enhanced $M_5$ but relatively unchanged $M_4$ intensity.
Therefore, this specific selective
excitation of Nb$M_5$ XES can be also used to receive
experimental information about Nb$5p$ DOS undistorted by overlapping
with the Nb $M_4$ XES.

\begin{figure}
\centerline{\epsfig{file=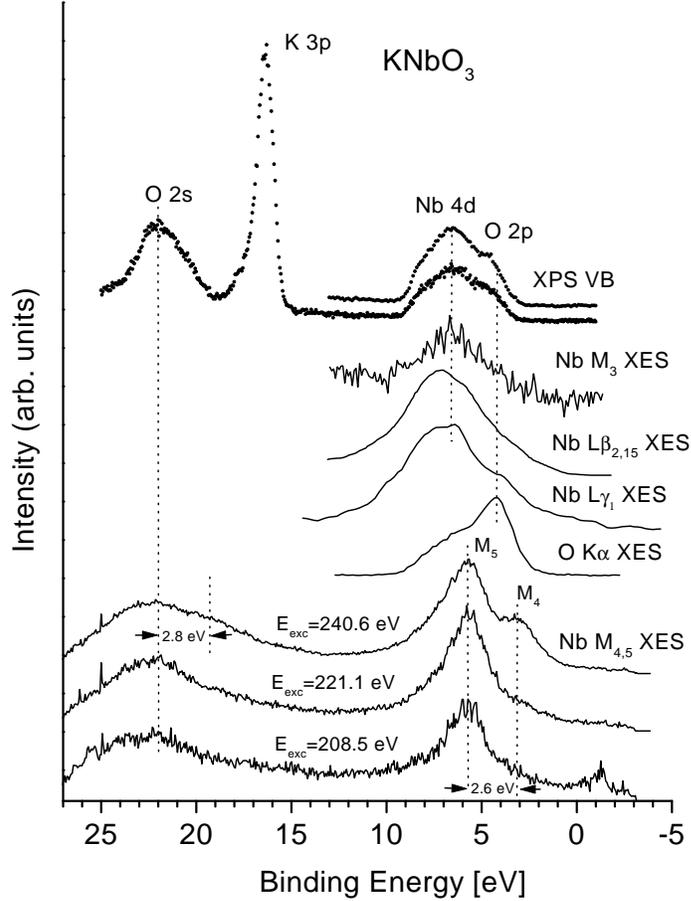,height=12.0cm}}
\caption{
The comparison of valence band photoemission (XPS VB),
O $K\alpha$-emission and several Nb XES in KNbO$_3$.
}
\label{fig:all}
\end{figure}

Apart from Nb $M_{4,5}$, some other XES measurements are useful
to provide complementary information about the electronic state
distribution at the Nb site. In Fig.~\ref{fig:all},
Nb $M_3$ ($4d{\rightarrow}3p_{3/2}$), Nb $L\beta_{2,15}$
($4d_{5/2,3/2}{\rightarrow}2p_{3/2}$) and
Nb $\gamma_1$ ($4d_{3/2}{\rightarrow}2p_{1/2}$) spectra are shown
in comparison with the VB XPS. The spectra are
brought to a common scale of binding energies, based
on the binding energies of corresponding core levels.
Since $M_4$ and $M_5$ spectra cannot be easily separated,
they are plotted with respect to the $3d_{5/2}$ binding energy (207.23 eV).
The Nb $M_{2,3}$ spectra have not been, to our knowledge, measured before,
since they were not listed in the Beardeen tables \cite{beardeen},
nor in the last systematic study of the ultrasoft XES
in $4d$-transition metals \cite{fomichev}.
These spectra lie in a convenient energy region for the
synchrotron study, and their two components ($3p_{1/2}$ and
$3p_{3/2}$) are well separated
by $\approx$15 eV. The $M_3$ spectrum
shown in Fig.~\ref{fig:all} was obtained in an exposition time
of about 20 min. Despite their low intensity, the $M_{2,3}$
spectra may be potentially suitable for the study of the VB,
probably including dispersion effects.
The overall difference between Nb$M_3$ and Nb$L$ spectra
on one hand and the Nb $M_{4,5}$ spectra on the other hand
is that the former probe primarily the occupied Nb$4d$ DOS,
centered near the VB bottom, whereas the latter
reveal the contribution of the occupied states with the $l\!=\!1$
symmetry at Nb site, that are diffuse and hybridize strongly
with VB states of other atoms. It is well seen
from Fig.~\ref{fig:all} that on the common energy scale,
the maximum of the $M_5$ spectrum lies roughly in the
middle of the VB. The energy positioning of the O $K_{\alpha}$ spectrum
in the common energy scale has been done based on the
O$1s$ binding energy (530.1 eV). This spectrum
reveals the predominant distribution of O$2p$ states near the top
of the VB -- the fact well known from band structure
calculations (see, for example, Ref.~\cite{ktn3}). The shape
of the spectrum shows no development
as the excitation energy varies, the reason being that no vacant
O$2p$ states are available near the bottom of the conduction band,
so that the O $K_{\alpha}$ emission is essentially uncoherent
with the corresponding absorption.
The $3p$ state of potassium does not hybridize with
any of the states contributing to the spectra above discussed
and is visible only in the XPS.

\begin{figure}
\centerline{\epsfig{file=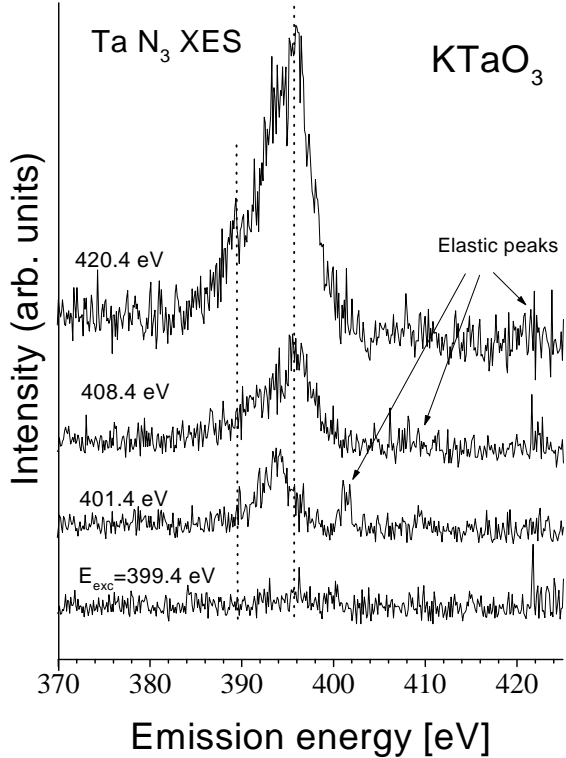,height=10.0cm}}
\caption{
Ta $N_3$ X-ray emission spectra of KTaO$_3$
for four excitation energies.
}
\label{fig:ktao3}
\end{figure}

Potassium tantalate has been earlier studied by XPS
in comparison with potassium niobate \cite{asa92,winiar}. In the present
work, we concentrate on the Ta $N_{2,3}$
($5d6s{\rightarrow}4p_{3/2,1/2}$) emission spectra
that have not been reported before.
The $M_2$ and $M_3$ spectra are far separated in energy
(by $\approx$62 eV) and probe primarily the Ta states
well represented in the VB and at the bottom of the
conduction band. As is known from the calculated band structure
(see Ref. \cite{ktn3,singh}, these
bands exhibit strong energy dispersion.
In Fig.~\ref{fig:ktao3},
the sequence of Ta $N_3$ is shown for several excitation energies,
obtained at 5--10 min. exposure per spectrum.
The intensity is therefore much higher
that that of counterpart $M_{2,3}$ spectra in KNbO$_3$.
Moreover, some development may be seen in the spectra
depending on the excitation energy.
While still absent at $E_{exc}$=399.4 eV, the X-ray emission
is clearly visible at $E_{exc}$=401.4 eV, i.e. well below
the Ta$4p_{3/2}$ excitation energy (404.0 eV). The displaced
peak position at $E_{exc}$=401.4 eV.
may be an indication of the X-ray resonant Raman scattering
\cite{raman} that enables excitations just below the threshold.
Similar trends have been observed in Ref.~\cite{uehara}
below and through the Ti $L_3$ absorption threshold
in another perovskite system (Ba,Sr)TiO$_3$.

The XPS of KTaO$_3$ does not differ much from earlier studies
\cite{asa92,winiar}. The O $K_{\alpha}$ emission spectrum is similar
(but not completely identical) to that in KNbO$_3$.
A separate analysis of our spectroscopy studies on KTaO$_3$
will be described elsewhere \cite{kto-new}.

Summarizing, we have compared the available spectroscopic
experimental information about
electronic structure of KNbO$_3$ (including new Nb $M_{4,5}$ XES,
Nb $L\beta_{2,15}$, Nb $L\gamma_1$, O $K_{\alpha}$ XES and
the VB XPS with the results of first-principles calculations.
The contributions from decay of the $3d_{5/2}$ and $3d_{3/2}$ holes
via valence emission are attributed by tuning the excitation energy.
In terms of the spectra being potentially suitable
for the study of band dispersion in the perovskites
in question, we found that Nb $M_{2,3}$ have very low intensity
whereas the two components of Nb $M_{4,5}$ are strongly overlapped
which complicates the analysis. The Ta $N_{2,3}$ spectra are
apparently free from both these disadvantages.
The implications for the structural studies in perovskites
may be -- provided the band dispersion studies turn out
to be successful -- the analysis of the band structure
distortion in lower-symmetry phases with the use
of angle-resolved resonance X-ray emission.

\section*{Acknowledgements}
This work was supported by the Russian Science Foundation
for Fundamental Research (Project 96-15-96598 and 98-02-04129),
the NATO Linkage Grant (HTECH.LG 971222), the DFG-RFFI Project,
the Swedish Natural Science Research Council (NFR) and the G\"oran
Gustavsson Foundation for Research in Natural Sciences and Medicine.
AVP, BS and MN acknowledge the support of the
German Research Society (SFB 225).


\begin{thebibliography}{99}
\bibitem{note}     The use of the density functional theory-based schemes
                   for the description of excited electronic states
                   has no justification but is widely used in practice.
                   From the experience based on quasiparticle calculations
                   available for some oxide systems, one can expect
                   an almost uniform energy shift accompanied by some slight
                   distortion of empty energy bands as a result of switching
                   from LDA to a quasiparticle calculation.
                   Moreover, the linearization error of the LAPW method
                   may become noticeable within the energy range
                   of $\approx$40 eV used in the present calculation
                   of the absorption spectra. One can expect, however,
                   that the hierarchy of excitation peaks is described
                   qualitatively correctly by the LDA in the vicinity
                   of the absorption threshold.
\end{thebibliography}
\end{document}